\documentclass[]{spie}  
\usepackage[utf8]{inputenc} 
\usepackage[T1]{fontenc}    
\usepackage{hyperref}       
\usepackage{url}            
\usepackage{booktabs}       
\usepackage{amsfonts}       
\usepackage{nicefrac}       
\usepackage{microtype}      
\usepackage{amsmath}

\usepackage{color}
\usepackage{amsmath,amssymb,bbm}
\usepackage{algorithm,algorithmicx}
\usepackage{algpseudocode}
\usepackage{enumerate}
\usepackage{graphicx}

\algnewcommand\algorithmicinput{\textbf{Init:}}
\algnewcommand\Init{\item[\algorithmicinput]}

\begin{document}

\title{Grounding Natural Language Commands to StarCraft II Game States for Narration-Guided Reinforcement Learning} 

\author{Nicholas Waytowich\supit{a}, Sean L. Barton\supit{b}, Vernon Lawhern\supit{a}, Ethan Stump\supit{b} and Garrett Warnell\supit{b}
\skiplinehalf
\supit{a}Human Research and Engineering Directorate, U.S. Army Research Laboratory \\
\supit{b}Computational and Information Sciences Directorate, U.S. Army Research Laboratory
}

\maketitle

\begin{abstract}
While deep reinforcement learning techniques have led to agents that are successfully able to learn to perform a number of tasks that had been previously unlearnable, these techniques are still susceptible to the longstanding problem of {\em reward sparsity}. 
This is especially true for tasks such as training an agent to play StarCraft II, a real-time strategy game where reward is only given at the end of a game which is usually very long.
While this problem can be addressed through reward shaping, such approaches typically require a human expert with specialized knowledge.
Inspired by the vision of enabling reward shaping through the more-accessible paradigm of natural-language narration, we investigate to what extent we can contextualize these narrations by grounding them to the goal-specific states.  We present a mutual-embedding model using a multi-input deep-neural network that projects a sequence of natural language commands into the same high-dimensional representation space as corresponding goal states.
We show that using this model we can learn an embedding space with separable and distinct clusters that accurately maps natural-language commands to corresponding game states . We also discuss how this model can allow for the use of narrations as a robust form of reward shaping to improve RL performance and efficiency.



\end{abstract}
\section{Introduction}
\label{sec:intro}

One of the chief goals in the field of artificial intelligence is to design agents that are capable of solving {\em sequential decision making} problems, i.e, problems in which an intelligent agent is expected not only to make predictions about the world, but also to act within it, and do this continuously over a certain period of time.
Of course, in order it to determine precisely {\em how} it should act, the agent must be provided with a clearly-defined goal.
Human designers typically communicate this information using a {\em reward function}, i.e., a scalar-valued rating of each state the agent may find itself in; positive reward values are defined for states the designer deems to be good, and negative reward values are used for those states deemed to be bad. 

Using a class of techniques called {\em reinforcement learning} (RL), an agent can, from its own experience, learn a {\em policy}, or specification of how it should act in order to accomplish its goals.
Importantly, the agent does not have access to the reward function itself, but rather only observes the reward values associated with the states it experiences.
Broadly, RL techniques perform learning by examining the observed reward values and modifying the agent's policy in order to favor repeating those actions that led to positive reward and avoiding those that led to negative reward.
Because the reward values play such a central role during learning, the efficacy of RL techniques is highly dependent on certain qualities of the specified reward function.

One quality of particular importance is referred to as {\em reward sparsity}, which is a measure of how many states in which the designer has specified nonzero (i.e., meaningful) reward values.
Reward functions with fewer nonzero values than others are said to be more sparse, and it is often easier for human designers to specify very sparse reward functions.
For example, if one were to design a reward function for solving StarCraft II (a complex real-time strategy (RTS) game with fast-paced actions and long time horizons), one could simply set the reward value to be a positive number when the agent has won the game, zero for all other positions/outcomes and therefore not have to think about how to define nonzero rewards elsewhere. 
However, sparse reward functions also negatively impact the efficacy of RL techniques.
Intuitively, this drop in efficacy comes about because the agent receives less meaningful feedback while attempting to perform its task; observed reward values of zero typically lead to no changes to the agent's policy.
In the StarCraft II example, the agent may spend a considerable amount of time taking random actions in the environment before it happens upon the goal state and receives any meaningful feedback that allows it to update it's policy. 


One class of methods that seeks to address this challenge of reward sparsity is that of {\em reward shaping}.
Reward shaping techniques allow one to modify the agent's reward function in order to encourage faster learning.
Unfortunately, many reward shaping paradigms require that the reward function be modified by humans that both have have a certain level of familiarity with how the agent was programmed and have the knowledge and access necessary to modify that programming.
In a vision of the future in which autonomous agents serve and team with humans of all sorts, we must enable paradigms of shaping that are accessible to people without this specialized knowledge.

We are motivated here by the vision of reward shaping through the accessible paradigm of narration.
By narration, we mean that human designers perform reward shaping not by modifying the source code of the agent, but rather by providing the agent with a sequence of natural-language commands that conveys the humans' suggestions for how that agent should go about accomplishing the task. 

In this paper, we seek to imbue the agent with an understanding of natural language commands by grounding those commands to goal-specific states in the environment. 
We do this by training a mutual-embedding model using a multi-input deep-neural network that projects a sequence of natural language commands into the same high-dimensional representation space as corresponding goal states.
Our goal is to eventually train a narration-guided reinforcement learning agent to play StarCraft II and as such we first focus on developing a mutual-embedding model that can handle the high dimensionality and complexity of StarCraft II game states. 
Specifically, we investigate to what extent natural language commands can be grounded to game states for allowing and agent to infer desired goals and intent. We Then discuss how this mutual embedding approach could be used for reward shaping in a reinforcement learning context.

\section{Related Work}
\subsection{Challenges for Reinforcement Learning}

The ability for RL to learn an optimal policy $\pi$ depends critically on how densely, or frequently, reward signals are provided from the environment. For example, it is possible to train an agent that can outperform human players for many Atari games when using the game score as a dense reward, yet for games which provide very sparse rewards (i.e.: Montezuma's Revenge), the learned policy was significantly worse than a human player \cite{Mnih2015a}. However, in many situations the reward signals are often very sparse, for example using a win/loss reward signal for solving games such as Go \cite{silver2016}, a game which has long time horizons. These domains require both significant data and computational resources to solve for the optimal policy. One potential approach to alleviate this issue is through \textit{reward shaping} \cite{dorigo1998robot,Mataric1997}, whereby the environment reward function is externally modified by, for example, a human observer in order to provide more frequent rewards and improve the stability and speed of policy learning. Human observers can provide these reward signal modifications in a multitude of ways, from using demonstrations of the task \cite{Argall2009, Goecks2018}, to binary good/bad feedback \cite{Knox2009, MacGlashan2017, Warnell2018} to natural language guidance \cite{Arumugam2017, Matuszek2013, Blukis18, Sung2018, Shah2018, MacGlashan-RSS-15}. In this work we focus on reward shaping using natural language guidance.

\subsection{Narration-Guided Reinforcement Learning}
There has been extensive prior work on using natural language based instruction from which to learn policies for autonomous agents across a variety of domains, from applications to text-based adventure games \cite{He2016ACL} to learning language-guided autonomous policies for robotic navigation and obstacle avoidance \cite{Mei2016AAAI,Arumugam2017,Matuszek2013,Blukis18,Sung2018,Shah2018, MacGlashan-RSS-15}. Recently, Bahdanau et. al. \cite{Bahdanau2018} developed AGILE (Adversarial Goal-Induced Learning from Examples), a system where a policy conditioned upon natural language instructions is trained from a discriminator which determines weather the the policy has reached a goal state, also conditioned on the current instruction. 
Another recent work by Fu et. al. \cite{Fu2018ICLR} proposed a language-conditioned reward learning (LC-RL) framework, which grounded language commands as a reward function represented by a deep neural network. They demonstrated that their model learned rewards that transfer to novel tasks and environments on realistic, high-dimensional visual environments with natural language commands. Tung et. al. \cite{Tung_2018_CVPR} collected a narrated visual demonstration (NVD) dataset where human teachers performed activities while describing them in detail. They then mapped the teachers’ descriptions to perceptual reward detectors which they then used to train corresponding behavioural policies in simulation. Kaplan et. al. \cite{Kaplan2017} applied a natural language reward shaping framework to the Atari game Montezuma's Revenge and showed that it outperformed existing RL approaches that did not use reward shaping.

\section{Methods and Approach}
\label{sec:meth}

\subsection{StarCraft II: BuildMarines Mini-game}
In this work we wish to investigate using natual-language commands to train an RL agent to play StarCraft II (SC2). Like most RTS games, StarCraft II is particularly challenging for RL algorithms to solve given the complicated rules, large action spaces, partial-observability of the environment state, long time-horizons and, most of all, due to the sparsity of the game score.  Because of these factors, our first goal was to solve one the simpler, yet still challenging, SC2 mini-games outlined in Vinyals et. al. \cite{Vinyals2017}. There are several mini-games defined, each with varying complexity and difficulty. In this paper, we focus on the most difficult mini-game called BuildMarines \cite{Vinyals2017}. The difficulty of this mini-game arises from its use of a very sparse reward function and as such it remains an open challenge for traditional state-of-the-art RL algorithms. 

As the name suggests, the main goal of this mini-game is to train an agent to build as many marines as possible in a certain time frame. To do this the agent must follow a sequential set of behaviors: build workers, collect resources, build supply depots, build barracks, and finally train marines. The agent starts with a single base and several workers that are automatically set to gather resources, and must learn to construct supply depots (which allow for more controller units to be built), as well as build marine barracks (which allow for marines to be generated) as interim steps before it can achieve its final goal. The agent receives a scalar-valued reward from the environment only when it successfully builds a marine, though it receives additional rewards for each marine built.  In this paper, we reduced the action space from that of the full StarCraft 2 action space to the minimum set of actions to reasonably accomplish the task (see below). 

\subsubsection{State and Action Spaces for BuildMarines Task}
We utilized the StarCraft II Learning Environment (SC2LE) API developed by DeepMind as the primary means for interacting with the StarCraft II (SC2) environment \cite{Vinyals2017}. Using SC2LE, the SC2 state-space consists of 7 mini-map feature layers of size 64x64 and 13 screen feature layer maps of size 64x64 for a total of 20 64x64 2d images (see left panel of Figure \ref{fig:stateInputProcessing}). Additionally, there are 13 non-spatial features that are also part of the state space containing information such as player resources and build queues.  The actions in SC2 are compound actions in the form of functions that require arguments and specifications about where that action is intended to take place on the screen. For example, an action such as ``build a  supply depot'' is represented as a function that would require the x-y location on the screen for the supply depot to be built. As such, the action space consists of the action identifier (i.e. which action to run), and an two spatial actions (x and y) that are represented as two vectors of length 64 real-valued entries between 0 and 1. 

\begin{figure}[!t]
	\begin{center}
		\includegraphics[width=0.99\linewidth]{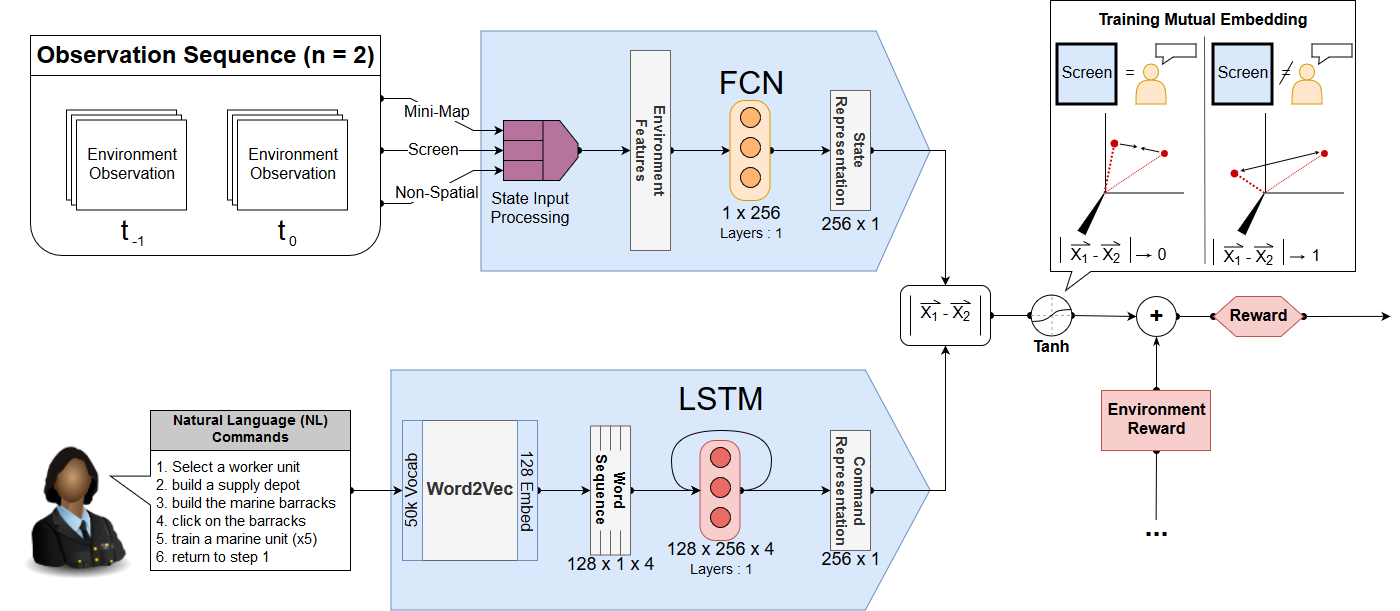}
		\caption{Diagram of the mutual-embedding model for deriving reward from natural language commands. The embedding model has two inputs: (Top) the StarCraft game states are fed through the state input processing module and a fully connected layer to transform into a 256-length state embedding. (Bottom) the corresponding natural language commands are passed through a word2vec encoder and then through an LSTM to end at a 256-length language embedding. The model then learns a common embedding space such that the euclidian distance of the state and command representations are pushed together if they correspond, or are pushed further apart otherwise (top right).}
		\label{fig:mutual-embedding}
	\end{center}
\end{figure}

\subsection{Grounding Natural Language Commands to Task States and Goals}
Our goal with this study was to overcome the problem of reward sparsity, wherein the success or failure of a task is determined only by the final state of the environment. Reward sparsity presents a critical problem for learning efficiency because agents must take a enormous number of random actions before stumbling upon the first instance of successful task completion. 
Here, we propose a mutual-embedding model to ground natural language commands to task states and goals. In doing this, we hope to facilitate a narration-guided approach to providing interim rewards to RL agents in the presence of sparse environment rewards.
\subsubsection{The Mutual-embedding Model}
In order for an RL agent to make use of narrations provided by a human, the language for those narrations needs to be grounded in a context that the agent can understand. 
We do this by training a \textit{mutual-embedding model} (MEM) that learns a contextual mapping between the StarCraft II game states and a set of natural language commands\footnote{The specific set of natural language commands that we use in our narration guided approach is shown on the bottom left of Figure \ref{fig:mutual-embedding}.} that indicate the desirable interim goals. The mutual-embedding model (shown in Figure \ref{fig:mutual-embedding} and discussed in detail below) learns a common representation of the game states and natural language commands that correspond to those game states. This common representation allows the agent to assign contextual meaning to the current game state that it is experiencing.

The model is learned by first projecting the language commands and game states into vector spaces of matching dimensionality, and then minimizing the euclidean distance between the vectors of corresponding commands and game states, while simultaneously maximizing the euclidean distance between commands and states that do not correspond. For example, we wish for game-state embeddings that correspond to the command "Build a supply depot" to be closer to that command's vector representation in the mutual embedding space, while being further away from all other command embeddings. The result is a shared embedding space that can represent both the semantic meaning of natural language commands, as well as the contextual meaning of game states. Ultimately, successfully training the MEM depends on three core processes: the embedding of the natural language commands, the embedding of the game-states, and the learned correspondence between these two embedding spaces that forms the MEM.

\textbf{\textit{Language Embedding:}}
To achieve a useful language embedding, we trained a word-level semantic mapping using word2vec \cite{Mikolov2013} with a vocabulary size of 50k and an embedding size of 128. Using this pre-trained word2vec model, word-level semantic embeddings were extracted from each word in a command and then passed through an LSTM model to create a command-level embedding of size 256 (bottom part of Figure \ref{fig:mutual-embedding}). The idea behind creating a command level embedding that is derived from word-level semantic embeddings is that it might allow for generalization to new commands composed of words with semantically similar meanings. For example, specific words in a command could be swapped out with semantically similar words (e.g., exchanging ``construct'' for ``build''), and the projection of the new command should be a near-neighbor to the projection of the original command.

\begin{figure}[!t]
	\begin{center}
		\includegraphics[width=0.99\linewidth]{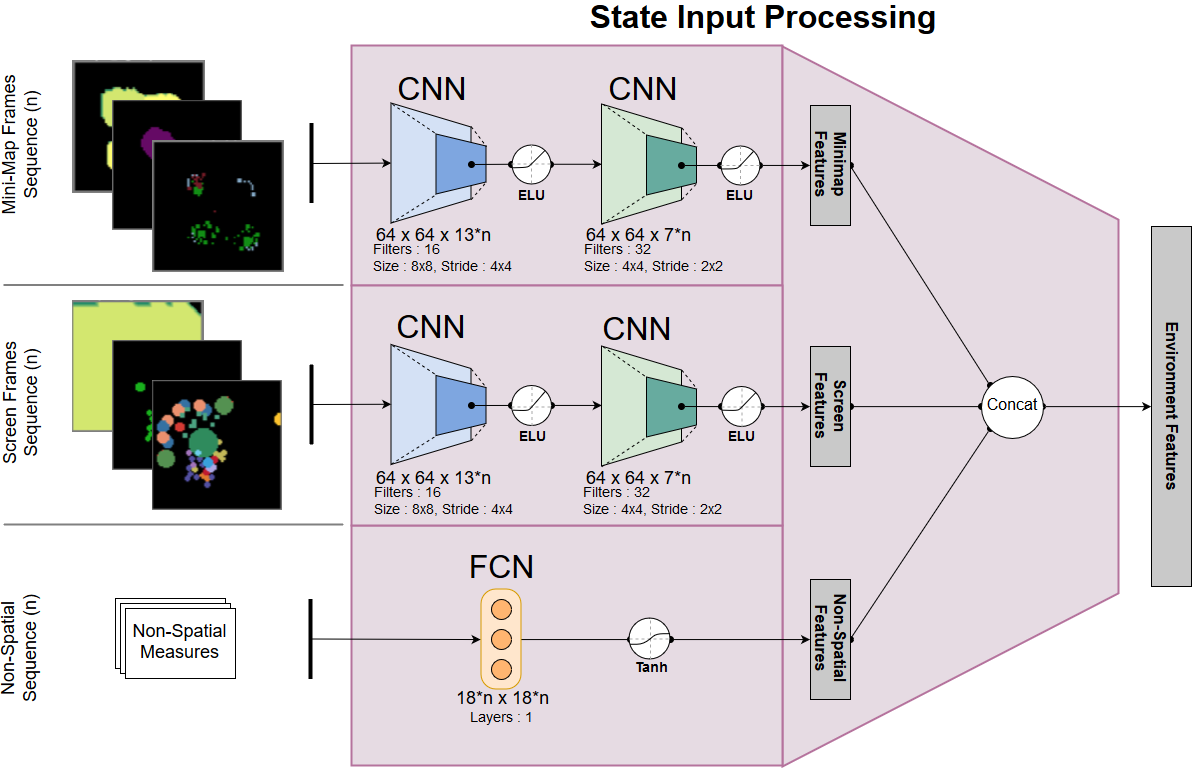}
		\caption{State input processing. Shown here is the state input processing pipeline for the mutual-embedding model and the RL agent for the SC2 task. SC2LE provides 3 primary streams of state information: mini-map layers, screen layers, and non-spatial features (such as resources, available actions and build queues). The mini-map and screen features were processed by identical 2-layer CNNs (top two rows) in order to extract visual feature representations of the global and local states of the map, respectively. The non-spatial features were processed through a fully-connected layer with a non-linear activation. These three outputs were then concatenated to form the full state-space representation for the agent, as well as for the state-based portion of the mutual-embedding model. In order to observe a a change in the state, a temporal stack (n=2) of inputs were processed during training.}
		\label{fig:stateInputProcessing}
	\end{center}
\end{figure}

\textbf{\textit{State Embedding:}}
With the command-level embedding defined, the second stage of the MEM is to project the game states of SC2 into a common embedding space. These game states (composed of mini-map, screen and non-spatial inputs from the SC2LE API) are processed by a state input processing module (shown in Figure \ref{fig:stateInputProcessing}, which consists of two branches of 2-d convolutional neural networks and fully connected network branch) as a feature extraction step. The SC2 screen and mini-map frames\footnote{A temporal stack of 2 frames (n=2) was used as the game-state input during training of the MEM in order to detect the change that corresponded to meeting the goal.} are each passed through a pipeline of two 2-d convolution layers that extract relevant feature vectors from the images. The non-spatial features are passed through a fully connected layers with a non-linear activation function to create a single non-spatial feature. The three feature outputs are then flattened and concatenated to produce a comprehensive feature array.
Finally, this comprehensive feature layer is projected into a 256-length embedding space using a final fully connected layer, thereby matching the dimensionality of the natural language command embedding.

\textbf{\textit{Mutual Embedding:}}
The mutual-embedding model itself (shown in Figure \ref{fig:mutual-embedding}) aims to capture a mutual representation of natural language commands and the corresponding game states that serve as interim goals. The model is trained such that game states are pushed closer to their corresponding language commands in the mutual embedding space and are pushed father away from non-corresponding commands. This is done by simultaneously training the embedding networks for both the game states and natural language commands to minimize the $\ell_2$-norm of the difference between the embedding vectors when the game state corresponds to the command, and maximizing the $\ell_2$-norm of the difference between the embedding vectors when the game state and commands do not correspond. The overall loss function used to train is shown below:
\begin{equation}
   \mathcal{L}(\theta) = \frac{1}{N} \sum_n^N \Big(||X_s - X_c|| - y\Big)^2 + \lambda||\theta||^2 
\end{equation}

\noindent where $\theta$ are the neural network embedding parameters, $||\cdot||$ corresponds to the $\ell_2$-norm, $\lambda$ is the $\ell_2$-norm penalty on the network weights ($\lambda = 2.5e^{-3}$ in our case), $X_f$ corresponds to the game state embedding, $X_c$ corresponds to the command embedding and $y \in \{0,1\}$ is the label representing if the game state and command are matching (congruent) or mismatching (incongruent). Our primary objective is to find $\hat \theta = \arg \min_{\theta} \mathcal{L}$, optimized over a set of $n = N$ training samples.

\textbf{\textit{Dataset Generation for Learning Mutual Embeddings:}}
To train the mutual-embedding between the natural language commands and the SC2 game states, a labeled dataset is needed that contains pairs of commands and states for supervised learning. In this case, game states corresponding to the different natural language commands need to be collected and there are two main ways this can be done. The first is by using a human (or some other agent) to play the game while following each instruction (or have a human watch the agent play the game) and save the game states when each instruction is reached. However, in order to train a mutual-embedding model consisting of a deep neural network, large numbers of examples are required and thus this option requires a significant amount human effort. The second approach is to use hand-crafted rules to automatically generate or label game states that correspond to each command. Although this option is less burdensome to collect the data, it requires the ability to hand-craft detectors which is not always possible in all tasks or situations. For this paper, we use the second approach to generate a data set since the SC2 state-space is rich enough to construct simple rules that correspond to each instruction. 

For each natural language command, we ran a random agent through the BuildMarines mini-game of the SC2 environment to generate a large set of game states. Then, we use hand-crafted rules that were set up to identify states that satisfy each command to generate corresponding labels. 
An example rule for labeling game states that satisfy the ``build a supply depot'' command, the number of screen pixels corresponding to a supply depot is tracked during the game play of the random agent and whenever that number increases (i.e., the agent has just build another supply depot), then instruction is considered satisfied and the corresponding state is labeled. 

We trained the mutual-embedding model (MEM) using a dataset consisting of pairs of game-states and command embeddings generated by the random agent.  This produced a dataset of 50k game-states consisting of 10k game state examples corresponding to each of the five language commands. We then created matching and mismatching pairs of labeled samples between the states and commands for a total of 100k labeled pairs (50k matched and 50k mismatched pairs). By training on mismatched pairs as well as matched pairs, the model learned not only to associate matched commands and states, but to strongly distinguish commands from other possible game states. Additionally, we included 50k ``null'' states which were game states that did not correspond to any of the language commands. This null set further distinguished desirable goal states from other states an agent might necessarily pass through. In total, we used a dataset size of 150k samples.

\textbf{\textit{Training the Mutual-Embedding Model}}
The dataset generated from the previous section was augmented to have congruent and in-congruent labels (i.e. matching and mismatching command/state pairs). The idea here is to learn a mutual embedding manifold where command embeddings are closer to the embeddings of the state they represent while simultaneously being further away from other states. The data were separated into training, validation and testing data and the mutual embeddings model was trained using an adam optimizer to minimize the loss function in Equation 1. The validation data was used to prevent model overfitting and training was stopped once the validation error was minimized. 
\section{Evaluation of the Mutual Embedding Model}
\label{sec:eval}
To evaluate performance, we split the dataset into three partitions where 100k samples were used for training, 25k samples were used for model validation, and 25k samples were held out for testing. We trained the MEM using an Adam optimizer with a learning rate of $5e^{-4}$ to minimize the loss function shown in Equation 1. The model was trained using a batch size of 32 over 20 epochs of the data. We chose the point of minimum validation loss to evaluate our model performance. We chose an embedding distance threshold of $0.5$ as the maximum distance required for a command embedding and game state embedding to be associated\footnote{A threshold is necessary to act as a decision boundary, and by using this threshold as an evaluation criterion we transform the mean squared error of the loss the into an accuracy.}. 
Our model was able to achieve a training accuracy of  $95.61$\% a validation accuracy of $82.35$\% and a test accuracy of  $80.40$\%. 


\subsection{Representation}

\begin{figure}[!t]
    \begin{center}
        \includegraphics[width=0.9\linewidth]{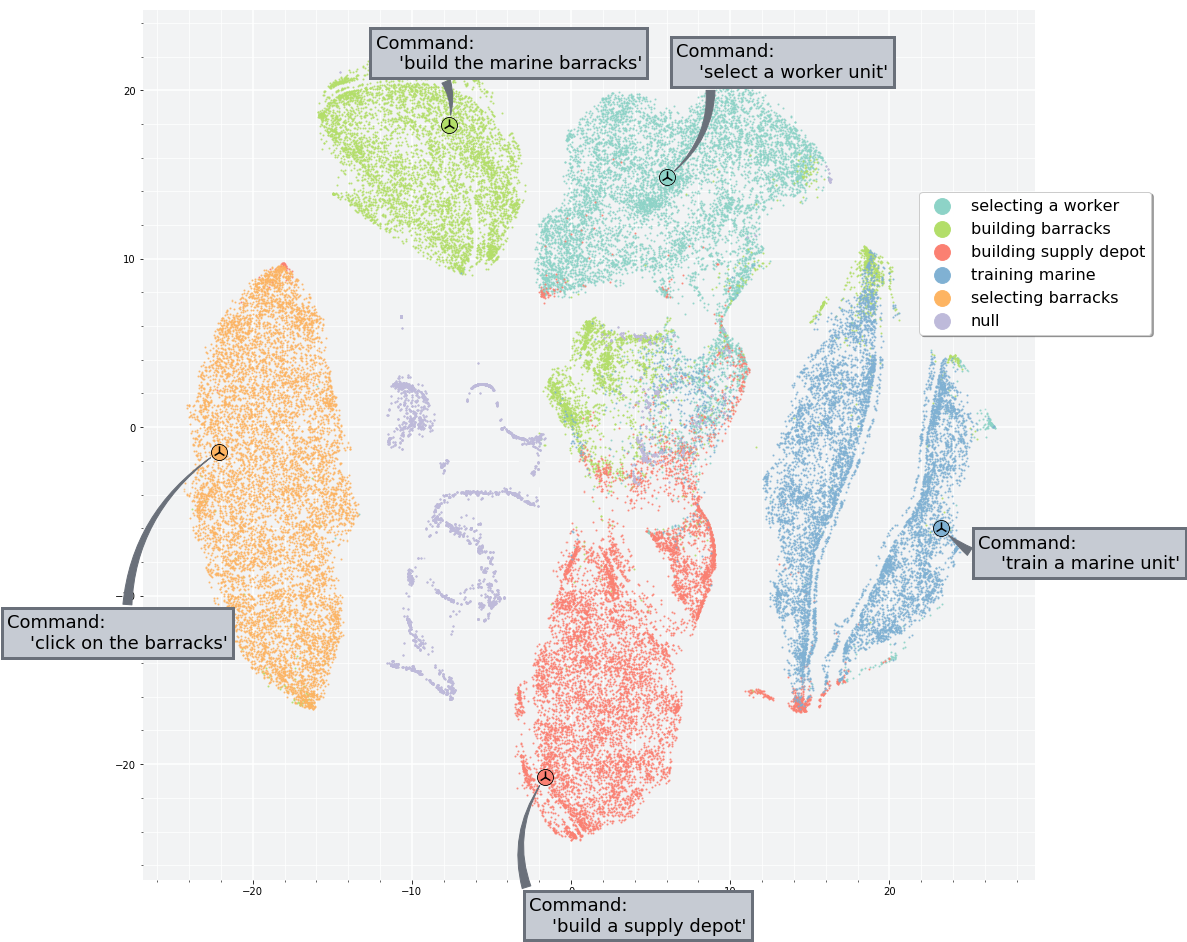}
        \caption{\footnotesize{Learned knowledge representation of the mutual-embedding model (MEM). Point clouds represent state examples, while the larger symbols with call-outs represent the command embedding projections. Here we used t-SNE to discover if a) the MEM could distinguish between relevant goal states, and b) if the MEM learned to project the natural-language commands into the same knowledge representation space as the state representations. t-SNE was implemented on the training data used to train the MEM. Separation of the state-space examples based on the goal state being demonstrated is clearly visible by the clustering of the data into distinct representation groups. Additionally, the natural-language commands that correspond with each goal state clearly project to the same manifold, demonstrating highly successful learning on the part of the mutual-embedding model.}}
        \label{fig:TSNE}
    \end{center}
\end{figure}

As discussed previously, the output of the mutual-embedding model is a pair of vector representations about the agents current state, as well as the objective represented with a natural language command (see Figure \ref{fig:mutual-embedding}). The success of the mutual-embedding in our design depends on the ability to project the natural-language commands in the same knowledge representation space as the state information provided by SC2, thereby recognizing from state information when a particular goal state has been reached. Thus, the MEM must learn two fundamental rules: how to distinguish between the possible goal states it could be in, and how to recognize if its current state matches the desired state provided by the natural language command. In order to assess the degree to which the MEM was able to learn these two judgements, we used t-distributed stochastic neighbor embedding (t-SNE) to visually inspect how the high-dimensional representations of the state and the commands were organized within the knowledge representation space of the neural network \cite{maaten2008visualizing}.

Because training of the MEM was a supervised learning task, we can easily inspect how states that correspond to examples of particular goals states (training a marine, building a supply depot, etc.) project into the representational space of the network. If the MEM has learned to successfully distinguish between goal states, then we should see clear separation of clusters and groups as a function of which state is being presented. Figure \ref{fig:TSNE} shows a clear separation between goal states in the MEM's network representation space. as evidenced by the distinct groups. While there are still some errors in classification visible (in particular near the boundary between ``build a baracks'' and ``build a supply depot'' which are highly similar events in the game), the ME model is clearly able to successfully distinguish between the critical goal states of the task, which is in agreement with the classification accuracies reported in the previous section.
Additionally, we projected the natural-language command representations into this same knowledge representation space in order to judge if they corresponded with the goal states they were meant to confer. Figure \ref{fig:TSNE} clearly shows that the high-dimensional representation of the commands clearly project to the manifolds that correspond with the learned goal states. 

\section{Discussion}
The work presented here investigated the ability for natural language commands to be embedded within the same space as high-dimensional tasks states, such as those used by an RL algorithm to develop an optimal policy. Specifically, we used the game StarCraft II as our task environment, as it provides and excellent bench mark for complex sequential RL with realistically difficult challenges. We presented a mutual embedding model that projects a sequence of natural language commands into the same high-dimensional representation space as corresponding goal states, such that these commands can be used as a robust form of reward shaping to improve RL performance and efficiency.

\subsection{Natural Language as Representation}
Humans routinely use natural language to communicate information about desired goals, plans, and strategies. Central to our ability to interact with one another is the understanding that the information communicated through language is grounded in realistic states of our world \cite{Harnad1990}. Because of this grounding, it is possible to know if the goals we communicate are met by comparing the representation postulated by language with our representation of the world, as defined by perceptual information. Here, we sought to model this type of mutual embedding by training a multi-input deep neural network to learn a mutual embedding between highly complex world states in the StarCraft II environment and natural language commands that were meant to indicate desirable sequential goals in a larger, more complex task. Using t-SNE, we are able to visualize the representation space learned by the MEM to represent relevant states of the game.

We also showed that natural language commands intended to communicate the desirable sub-goals are able to be projected into the same representation space occupied by the game states, and correspond to their paired situation. Thus, the MEM represents the command "build a supply depot" in the same high-dimensional space \textit{and} location as the actual game state of building a supply depot. From the model's perspective, this natural language command is grounded in the real-world execution of the command. This is critical, as it suggests that the MEM can ground natural language goals in an agent's state space and thus is capable of indicating desired goal states through natural language commands.

\subsection{Implications for Reinforcement Learning}
There is a growing interest in applying reinforcement learning approaches to multi-agent real-time strategy games, such as StarCraft II, because they represent complex tasks with realistic challenges and characteristics. Pang et. al. \cite{Pang2018} used a hierarchical reinforcement learning (HRL) approach, whereby a \textit{controller} policy is trained to (1) select a sub-policy from a collection of base policies and (2) select the time horizon for which to execute that sub-policy. Typically the sub-policies represent temporal actions which can solve particular sub-goals of the task (for example, sub-policies to build units and harvest resources), whereas the goal of the controller policy is to find an optimal sequence of sub-policies to solve the main objective (defeating the opponent). They found that a hand-crafted reward signal performed better than using a binary reward of 1/0 for win/loss. They also used curriculum learning against varying difficulties of the built-in game AI to gradually train their policy. By combining these different approaches their policies were able to obtain a $93\%$ win rate against the expert game AI when trained on a computer with 48 cores and 8 Tesla K40 GPUs for two days.

While narration-guided RL has shown promise in addressing complex RL problems, to our knowledge this is the first attempt to deploy a natural-language-based reward system in the StarCraft II environment. Specifically, we show here that natural-language commands can be grounded in the game states of StarCraft II, in spite of the complexity and dimensionality of these game states. As such, our mutual-embedding model provides a promising mechanism for creating a generalized sequential reward that capitalizes on a human's capacity to utilize higher order knowledge to achieve long-term goals. By providing a means for a human to guide a learning agent via natural language, generalizable sequential policies may be learned without the overhead of creating hand-crafted sub-tasks or checkpoints that would depend critically on expert knowledge about RL reward functions.

A promising consequence of using a natural-language based approach to guiding RL learning is that the inherent flexibility in natural language allows for flexibility in how reward-shaping is enacted. Semantically similar commands issued to an RL agent by way of the MEM should lead to similar goal-based rewards, even if the specific natural language commands have not previously been observed or learned. For example, swapping words like "build" for the semantically similar word "construct" should not lead to any appreciable difference in the sub-goal implicated by the command. This flexibility may make it considerably easier to guide RL without domain specific expertise.

Additionally, there is good reason to believe that using a natural-language based reward mechanism could lead to superior RL performance and efficiency. Because the natural-language commands are conditioned on the states of the game themselves, the goals such commands represent exist within the same representation space that an agent's policy is conditioned on. This one-to-one mapping between game-state representation and goal-state representation provides learning agents with a means of directly comparing their current state to the desired state. By contrast, sub-task-based intermediate rewards (such as those used in Pang et al.) are based on hand tuned features that are not guaranteed to be represented within an agent's neural network. As such, using a mechanism like the MEM proposed here provides a more reliable means of communicating seqeuntial tasks to a learning agent.



\bibliographystyle{spiebib}
\bibliography{gaw_nrw}

\end{document}